\documentclass{aa}

\usepackage{graphicx}
\usepackage{txfonts}
\usepackage{hyperref}
\usepackage{xcolor}
\usepackage[normalem]{ulem}
\graphicspath{ {./figures/} }

\begin{document} 

   \title{Alfv\'enic versus non-Alfv\'enic turbulence in the inner heliosphere as observed by Parker Solar Probe}


   \author{Chen Shi
          \inst{1}
          \and
          Marco Velli\inst{1}
          \and 
          Olga Panasenco\inst{2}
          \and 
          Anna Tenerani\inst{3}
          \and 
          Victor R\'eville\inst{4}
          \and 
          Stuart D. Bale\inst{5,6,7,8}
          \and 
          Justin Kasper\inst{9}
          \and 
          Kelly Korreck\inst{10}
          \and 
          J. W. Bonnell\inst{6}
          \and 
          Thierry Dudok de Wit\inst{11}
          \and
          David M. Malaspina\inst{12}
          \and
          Keith Goetz\inst{13}
          \and 
          Peter R. Harvey\inst{6}
          \and 
          Robert J. MacDowall \inst{14}
          \and 
          Marc Pulupa\inst{6}
          \and 
          Anthony W. Case\inst{10}
          \and 
          Davin Larson\inst{6}
          \and 
          J. L. Verniero\inst{6}
          \and 
          Roberto Livi\inst{6}
          \and 
          Michael Stevens\inst{10}
          \and 
          Phyllis Whittlesey\inst{6}
          \and 
          Milan Maksimovic\inst{15}
          \and 
          Michel Moncuquet\inst{15}
          }

   \institute{Earth, Planetary, and Space Sciences, University of California, Los Angeles, Los Angles, California, USA \\
    \email{cshi1993@ucla.edu} 
    \and
    Advanced Heliophysics 
    \and 
    Department of Physics, University of Texas at Austin, Austin, Texas, USA \\
    \email{anna.tenerani@austin.utexas.edu} 
    \and 
    IRAP, Universit\'e Toulouse III—Paul Sabatier, CNRS, CNES, Toulouse, France 
    \and 
    Physics Department, University of California, Berkeley, CA 94720-7300, USA 
    \and 
    Space Sciences Laboratory, University of California, Berkeley, CA 94720-7450, USA 
    \and 
    The Blackett Laboratory, Imperial College London, London, SW7 2AZ, UK 
    \and 
    School of Physics and Astronomy, Queen Mary University of London, London E1 4NS, UK 
    \and 
    University of Michigan, Ann Arbor, MI, USA 
    \and 
    Smithsonian Astrophysical Observatory, Cambridge, MA, USA 
    \and 
    LPC2E, CNRS and University of Orl\'eans, Orl\'eans, France 
    \and 
    Laboratory for Atmospheric and Space Physics, University of Colorado, Boulder, Colorado, USA 
    \and
    School of Physics and Astronomy, University of Minnesota, Minneapolis, MN 55455, USA 
    \and 
    Solar System Exploration Division, NASA Goddard Space Flight Center, Greenbelt, MD 20771, USA 
    \and 
    LESIA, Observatoire de Paris, Universit\'e PSL, CNRS, Sorbonne Universit\'e, Universit\'e de Paris, 5 place Jules Janssen, F-92195 Meudon, France 
    }

    \titlerunning{Alfv\'enic vs. non-Alf\'enic turbulence in the inner heliosphere}
    \authorrunning{Shi et al.}

   \date{}

 
  \abstract
   {Parker Solar Probe (PSP) measures the magnetic field and plasma parameters of the solar wind at unprecedentedly close distances to the Sun. These data provide great opportunities to study the early-stage evolution of magnetohydrodynamic (MHD) turbulence in the solar wind.}
   {In this study, we make use of the PSP data to explore the nature of solar wind turbulence focusing on the Alfv\'enic character and power spectra of the fluctuations and their dependence on the distance and context (i.e., large-scale solar wind properties), aiming to understand the role that different effects such as source properties, solar wind expansion, and stream interaction might play in determining the turbulent state.}
   {We carried out a statistical survey of the data from the first five orbits of PSP with a focus on how the fluctuation properties at the large MHD scales vary with different solar wind streams and the distance from the Sun. A more in-depth analysis from several selected periods is also presented.}
   {Our results show that as fluctuations are transported outward by the solar wind, the magnetic field spectrum steepens while the shape of the velocity spectrum remains unchanged. The steepening process is controlled by the ``age'' of the turbulence, which is determined by the wind speed together with the radial distance. Statistically, faster solar wind has higher ``Alfv\'enicity,'' with a more dominant outward propagating wave component and more balanced magnetic and kinetic energies. The outward wave dominance gradually weakens with radial distance, while the excess of magnetic energy is found to be stronger as we move closer toward the Sun. We show that the turbulence properties can significantly vary from stream to stream even if these streams are of a similar speed, indicating very different origins of these streams. Especially, the slow wind that originates near the polar coronal holes has much lower Alfv\'enicity compared with the slow wind that originates from the active regions and pseudostreamers. We show that structures such as heliospheric current sheets and velocity shears can play an important role in modifying the properties of the turbulence.}
   {}


   \maketitle
%

\section{Introduction}

Turbulence is a ubiquitous phenomenon in nature, arising in neutral fluids such as in Earth's atmosphere and ocean as well as in astrophysical plasmas. The study of plasma turbulence is of great necessity because it is deeply related to fundamental nonlinear plasma physics and is crucial in understanding various important processes in astrophysics, such as the heating and acceleration of the solar wind and the acceleration of high-energy particles.

In the solar wind, direct measurements have shown that fluctuations in the velocity and magnetic field display properties of well-developed turbulence \citep[e.g.,][]{Coleman1968}. One important feature of these fluctuations that appears to be contradictory with a well-developed turbulence is the high Alfv\'enicity, that is to say the strong correlation between velocity and magnetic field fluctuations invariably displaying the properties of Alfv\'en waves propagating away from the Sun \citep[e.g.,][]{BelcherandDavis1971}, even though
the solar wind is propagating much faster than any wave speed and should therefore advect fluctuations outward irrespective of the direction of their propagation. This Alfv\'enic turbulence is most prevalent in high-speed solar wind streams, and the Alfv\'enic property appears to decay with distance from the Sun and survive out to distances much greater than 1 AU only in the polar heliosphere at solar minimum \citep{bavassano1998cross}. Alfv\'enic turbulence is also nearly incompressible. Radial evolution of the power spectra and other quantities, such as cross-helicity (defined below), seems to confirm ongoing nonlinear dynamics \citep{Bavassanoetal1982,bavassano1998cross}, which for incompressible Alfv\'enic turbulence requires the interaction between colliding counter-propagating wave packets \citep{Iroshnikov1964,Kraichnan1965}.

\begin{figure*}
    \centering
    \includegraphics[angle=-90,width=0.8\textwidth]{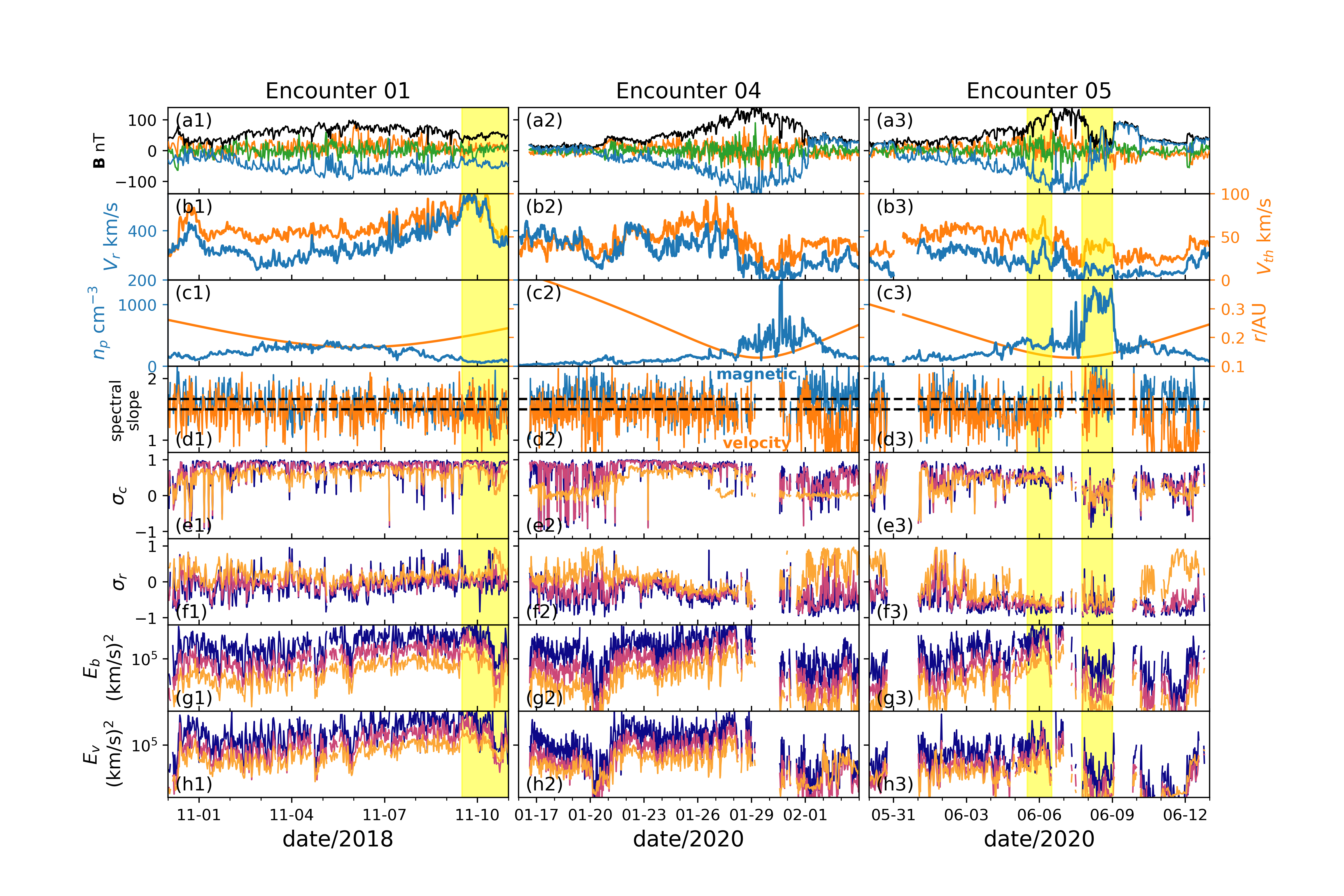}
    \caption{Overview of Encounter 01, 04, and 05. Row (a) shows the magnetic field with blue, orange, and green curves being radial, tangential, and normal components (RTN coordinates) and the black curve being the magnitude. Row (b) shows the radial ion flow speed (blue) and the ion thermal speed (orange). Row (c) shows the ion density (blue) and radial distance of PSP to the Sun (orange). Row (d) shows the spectral slopes of the magnetic field in Alfv\'en speed (blue) and velocity (orange). The two dashed lines mark the values $3/2$ and $5/3$ for reference. Row (e)-(h) are $\sigma_c$, $\sigma_r$, $E_b$, and $E_v$, respectively, as defined in Section \ref{sec:inst_data}. In each panel of these four rows, three curves are plotted and they correspond to wave band 2 (blue), 5 (purple), and 8 (yellow), respectively. All quantities were averaged or calculated through Fourier analysis in the $2048\times 0.874 \approx 30$min time window as described in Section \ref{sec:inst_data}.}
    \label{fig:overview}
\end{figure*}

Much theoretical work has been devoted to understanding the nature of nonlinear interactions in incompressible magnetohydrodynamic (MHD) turbulence. The early statistically isotropic phenomenological models of \citet{Iroshnikov1964} and \citet{Kraichnan1965} were extended to include parallel and perpendicular wave-number anisotropy by \citet{Goldreich1995}; the concept of dynamical alignment \citep{dobrowolny1980properties} was introduced to explain the dominance of outwardly propagating Alfv\'enic turbulence in the solar wind as a nonlinear phenomenon, and this was shown to lead to different spectra for inward and outward fluctuations by \citet{Grappinetal1990}.  A phenomenology of anisotropic turbulence with a preferred sense of propagation was presented in \citet{lithwick2003imbalanced}. 

The above models predict different spectral slopes and energy cascade rates, each under specific assumptions. However, solar wind observations cannot be solely explained by any one of the models, and this may be due to the inapplicability of assumptions such as homogeneity and incompressibility: The wind expands spherically, slowing nonlinear interactions and providing a quasi-scale free energy loss in the turbulence; different velocity streams with significant shear are present at meso-scales; and compressible processes such as parametric decay may occur \citep{primavera2019parametric,tenerani2020magnetic}, not to mention the potential role of particle distribution function anisotropies.

Two important problems that stand out are the different spectral slopes for the magnetic field and velocity \citep[e.g.,][]{Grappinetal1991,boldyrev2011spectral,Chenetal2013} and the observed excess of magnetic energy over the kinetic energy \citep[e.g.,][]{roberts1987nature,MarschandTu1990,Grappinetal1991}. It is observed beyond 0.3AU that the velocity spectrum, whose slope is around $-3/2$, is shallower than the magnetic field spectrum, whose slope is close to $-5/3$. Evidence shows that beyond 1 AU, the velocity spectrum steepens toward a $-5/3$ slope, implying an active nonlinear process \citep{roberts2007agu,bruno2013solar}. This nonlinear process, however, is not captured by the Alfv\'enic turbulence models. The observed magnetic energy excess is potentially a natural result of MHD turbulence evolution \citep[e.g.,][]{grappin1983dependence,boldyrev2009spectrum}, but it may also be explained by the convective magnetic structures in the solar wind \citep{tu1991case,tu1993model}. 

As mentioned before, the solar wind, instead of being a homogeneous medium, is radially stratified due to the spherical expansion. This inhomogeneity linearly couples the outward and inward propagating waves through the reflection of them \citep[e.g.,][]{Vellietal1991,Velli1993}. In addition, the radial expansion generates a new anisotropy with respect to the radial direction which mixes with the anisotropy with respect to the background magnetic field direction \citep{GrappinandVelli1996,dong2014evolution,TeneraniandVelli2017,Shietal2020}.

Parker Solar Probe (PSP), launched on August 12, 2018, has completed its first five orbits, with a closest approach of $\sim 27.9$ solar radii ($R_s$) to the Sun in encounters (E) E4 and E5, which is much closer than the previous record held by Helios B at $\sim 62.4$ $R_s$. Thus, its data provide a unique opportunity to study solar wind turbulence  in its early stage of evolution. Initial PSP data have revealed many interesting phenomena, among which the omnipresence of the so-called magnetic switchbacks may be especially
important \citep{Baleetal2019,deWitetal2020,McManusetal2020,tenerani2020magnetic}. There are fluctuations in the solar magnetic field of a sufficient magnitude to invert the local direction with respect to the Sun, that is to say they switch the field backward locally into a fold. Intriguingly, these folds retain some features typical of  Alfv\'enic turbulence, among which the strong correlation of velocity to magnetic field fluctuations as well as a nearly constant magnitude of the total magnetic field. The velocity-magnetic field correlation and the outward sense of propagation from the Sun reveal themselves through the presence of radial velocity outward jets superimposed on the background solar wind flow \citep{matteini2014dependence,kasper2019alfvenic,horbury2020sharp}.

\citet{reville2020role}, via MHD simulations  compared to the PSP data, show that the Alfv\'enic fluctuations provide sufficient power to accelerate the measured slow solar wind streams. \citet{chen2020evolution} surveyed the PSP data from the first two orbits and analyzed the Alfv\'enicity of the MHD turbulence in the solar wind. They show that the dominance of the outward-propagating wave decreases with radial distance to the Sun, which is consistent with previous observations made beyond 0.3 AU \citep{roberts1987origin,bavassano1998cross}. In addition, a steepening of the magnetic field spectrum  from a slope around $-1.5$ toward $-1.67$ is also observed. 

In this study, we make use of the PSP data from the first five orbits and conduct a statistical analysis of the MHD fluctuations in the solar wind. We show how the properties of the turbulence vary with both radial distance to the Sun and the wind speed. The wind speed in combination with the radial distance controls the turbulence spectra via the useful concept of turbulence ``age'' \citep{Grappinetal1991}. The Alfv\'enicity has a complicated behavior. In general, the fast wind is more Alfv\'enic than the slow wind and the Alfv\'enicity, if defined by the relative amplitude of the outward and inward propagating Alfv\'en waves, gradually decreases with radial distance. However, the magnetic energy seems to be much larger than the kinetic energy close to the Sun and gradually relaxes to similar levels as the wind propagates. In addition, Alfv\'enicity of streams of a similar speed can be very different. We discuss several factors that possibly influence the turbulence properties, including fast-slow stream shears, the heliospheric current sheet, and the different origin of the solar wind streams at the Sun.

\section{Instruments \& data processing}\label{sec:inst_data}
There are four instrument suites onboard PSP. Here we make use of the Level-2 magnetometer (MAG) data from Fields Experiment (FIELDS) and Level-3 Solar Probe Cup (SPC) data from Solar Wind Electrons Alphas and Protons investigation (SWEAP).  We refer to the five orbits of PSP as ``Encounter 1, 2, 3, 4, 5,'' respectively, or ``E 1, 2, 3, 4, 5'' for short, as high resolution data are only produced near perihelions of the orbits ($R \leq 0.3-0.4$ AU). During the encounters, SPC measures the proton spectrum at a cadence of 0.218-0.874s and the time resolution of FIELDS is smaller than 13.7ms. The exact time periods that are analyzed in this study are listed in Table \ref{tab:time_periods}.

\begin{table*}[t]
\caption{Time periods selected for analysis of the PSP data. The third column shows PSP perihelion dates and the fourth column shows the distance of each perihelion to the Sun.} 
\label{tab:time_periods}      
\centering              
\begin{tabular}{c c c c}        
\hline               
Encounter \# & Period & Perihelion date & Perihelion to the Sun\\    
\hline                        
  1 & Oct 31-Nov 11, 2018 & Nov 06, 2018 & 35.7$R_s$\\      
  2 & Mar 30-Apr 11, 2019 & Apr 04, 2019 & 35.7$R_s$\\ 
  3 & Aug 22-Aug 31, 2019 & Sep 01, 2019 & 35.7$R_s$\\ 
  4 & Jan 16-Feb 04, 2020 & Jan 29, 2020 & 27.9$R_s$\\ 
  5 & May 30-Jun 13, 2020 & Jun 07, 2020 & 27.9$R_s$\\ 
\hline                                   
\end{tabular}
\end{table*}

We first resampled the measurements of the magnetic field, proton density, velocity, and thermal speed to a time resolution of 0.874s, which is enough for the purpose of analyzing MHD turbulence. Then we binned the data into 2048-point time windows and filled the data gaps using linear interpolation. Windows with a data gap ratio larger than 10\% were discarded. We determined the polarity of the radial magnetic field by averaging $B_r$ inside each time window and defined the two Els\"asser variables
\begin{equation}\label{eq:definition_Zoi}
    \mathbf{Z_{o,i}} = \mathbf{U} \mp \mathrm{sign}(B_{r0}) \frac{\mathbf{B}}{\sqrt{\mu_0 \rho}}
\end{equation}
where subscripts ``o'' and ``i'' represent ``outward'' and ``inward,'' respectively, and $B_{r0}$ is the averaged $B_r$. We note that to have well-defined outward and inward propagating waves, the angle between the background magnetic field and the radial direction should not be too large. One can estimate that for a solar wind speed of 300 km/s, the spiral angle of the magnetic field is approximately 20 degrees at 60 solar radii, which is sufficiently small. In Eq (\ref{eq:definition_Zoi}), the density is the averaged value in each half-hour window. As is shown in Fig. \ref{fig:statistic_n_fluc_T_vr}, the relative density fluctuation $\Delta n /n$ is mostly small with values around 0.05-0.10. This density fluctuation introduces a small, negligible uncertainty, around $(2.5-5)\%$ when calculating the Alfv\'en speed.

Fourier transforms were applied to $\mathbf{U}$, $\mathbf{V_A} = \mathbf{B}/\sqrt{\mu_0 \rho}$, and $\mathbf{Z}_{o,i}$ to obtain power spectra. We then fit the power spectra over modes 5-60, which correspond to periods $T \in [30s,360s]$, which are within the inertial range of the turbulence.
Similar to \citet{Grappinetal1991}, we divided the Fourier modes into ten logarithmic bands, such that band $i$ includes modes $[2^{i-1},2^i)$. Inside each band, integrated wave energies $E_{b}, E_v, E_o,$ and $E_i$ were calculated. Then we defined the normalized cross helicity
\begin{equation}
    \sigma_c = \frac{E_o - E_i}{E_o + E_i}
,\end{equation}
which measures the relative amplitude of outward and inward Alfv\'en wave energies, and the normalized residual energy
\begin{equation}
    \sigma_r = \frac{E_v - E_b}{E_v + E_b}
,\end{equation}
which measures the relative amplitude of kinetic and magnetic energies. We note that $\sigma_c = \pm 1$ corresponds to purely outward and inward propagating Alfv\'enic fluctuations, while $\sigma_r = \pm 1$ corresponds to fluctuations that are either purely in the velocity field (kinetic) or magnetic. For purely outwardly-propagating Alfv\'enic fluctuations, we expect $\sigma_c=1$ and $\sigma_r=0$.

\begin{figure}
    \centering
    \includegraphics[width=\hsize]{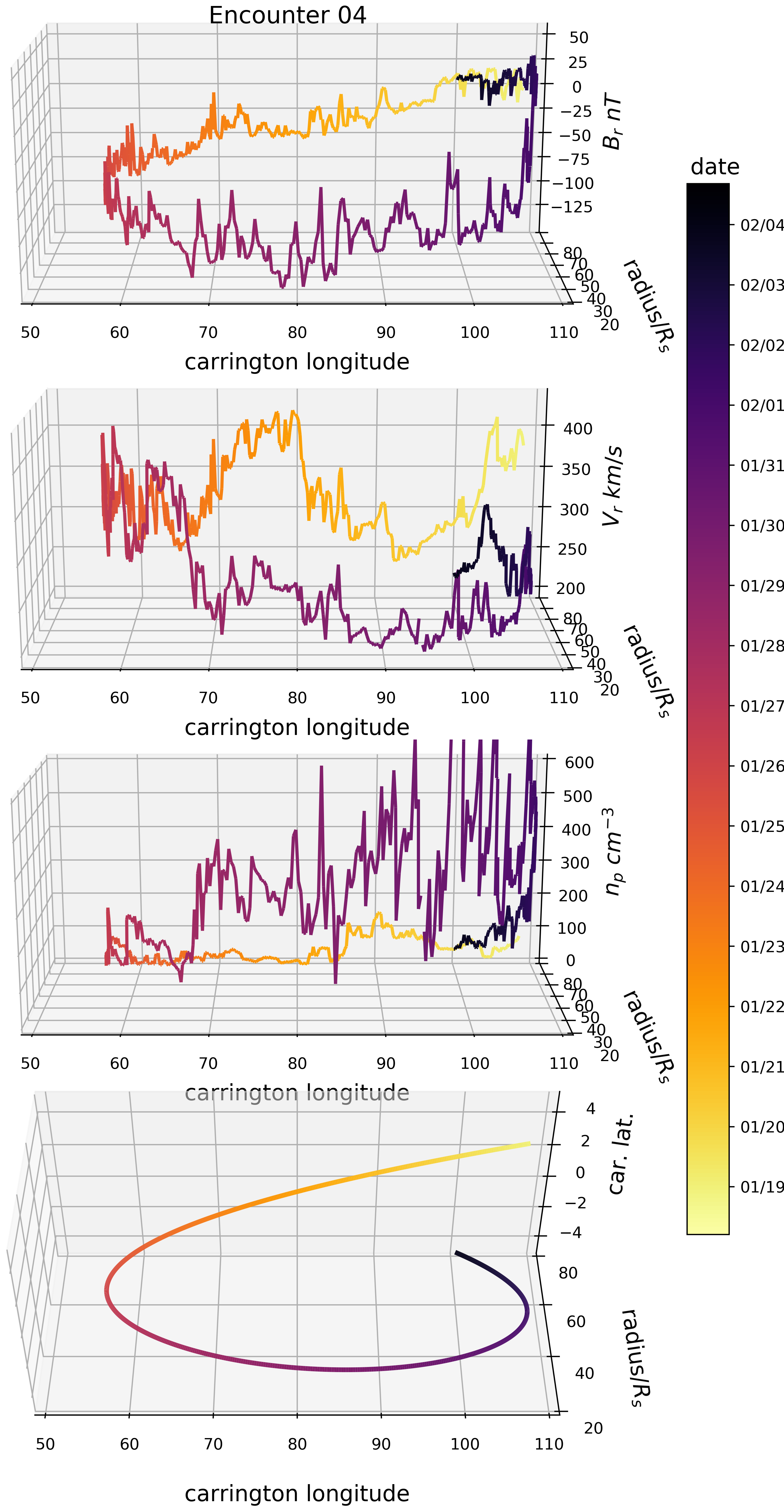}
    \caption{Measurements of radial magnetic field (top panel), radial flow speed (second panel), and proton number density (third panel) during Encounter 4. The values were plotted on a radius-(Carrington longitude) grid, i.e., in the reference frame corotating with the Sun. The bottom panel shows PSP's orbit with the $z$-axis being the Carrington latitude. We note that the variation in latitude is small. The colors represent time such that PSP moves from the light-colored end to the dark-colored end.}
    \label{fig:3D_trajectory_E04}
\end{figure}
\begin{figure}
    \centering
    \includegraphics[width=\hsize]{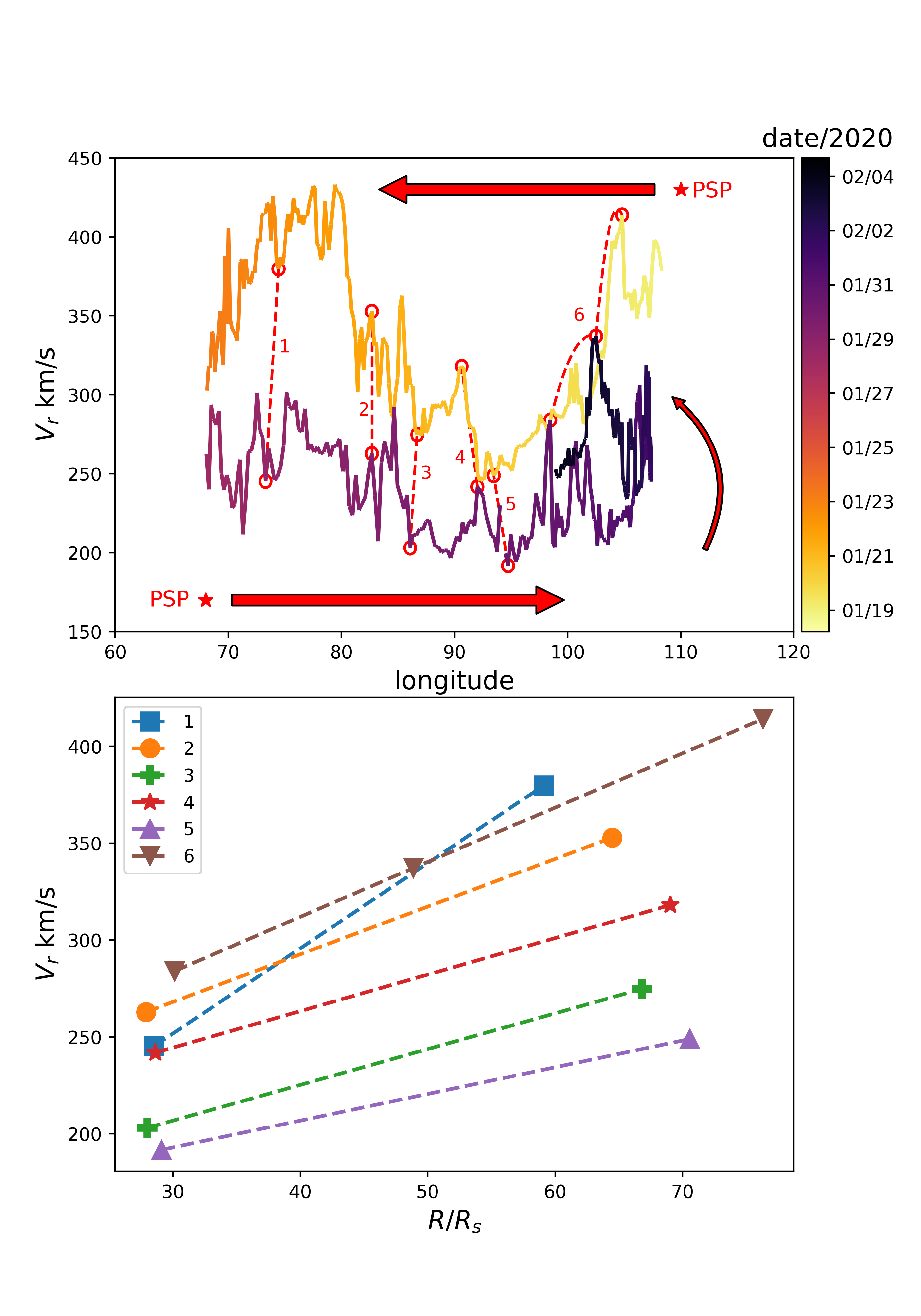}
    \caption{Top: Radial solar wind speed varying with Carrington longitude of PSP measured during Encounter 4. The colors represent the time and PSP travels from the light-colored end toward the dark-colored end as indicated by the red arrows. Red circles connected by dashed lines mark several selected structures that were observed at different radial distances. Bottom: Radial wind speed as a function of radial distance to the Sun. Each line corresponds to one single structure marked by the connected red circles in the top panel. The structures are numbered as 1, 2, 3, 4, 5, and 6, as annotated in the top panel.}
    \label{fig:Vr_carr_long_E04}
\end{figure}

\section{Results}
\subsection{Overview of Encounter 1, 4, \& 5}\label{sec:overview}
In Fig. \ref{fig:overview} we present the overview plot of Encounter 1 (left), 4 (middle), and 5 (right). We did not plot Encounter 2 \& 3 due to the limited figure size and less data coverage during the two encounters. All quantities were calculated, either averaged or Fourier-analyzed, in the $2048\times 0.874 \mathrm{s} \approx 30$ min time window as described in Section \ref{sec:inst_data}. Consequently, the magnetic switchbacks, whose typical time scales are several minutes long \citep{deWitetal2020}, are absent from the magnetic field plot. We note that the large gaps in the last four rows ($\sigma_c$, $\sigma_r$, $E_b$, and $E_v$) of the middle and right columns do not mean that the original SPC and MAG data have large gaps. The reason is, as mentioned in Section \ref{sec:inst_data}, that we discarded the half-hour time windows with data gap ratios larger than 10\%. So these large gaps are actually a result of frequent small data gaps in Encounters 4 \& 5.

\begin{figure*}
    \centering
    \includegraphics[width=\textwidth]{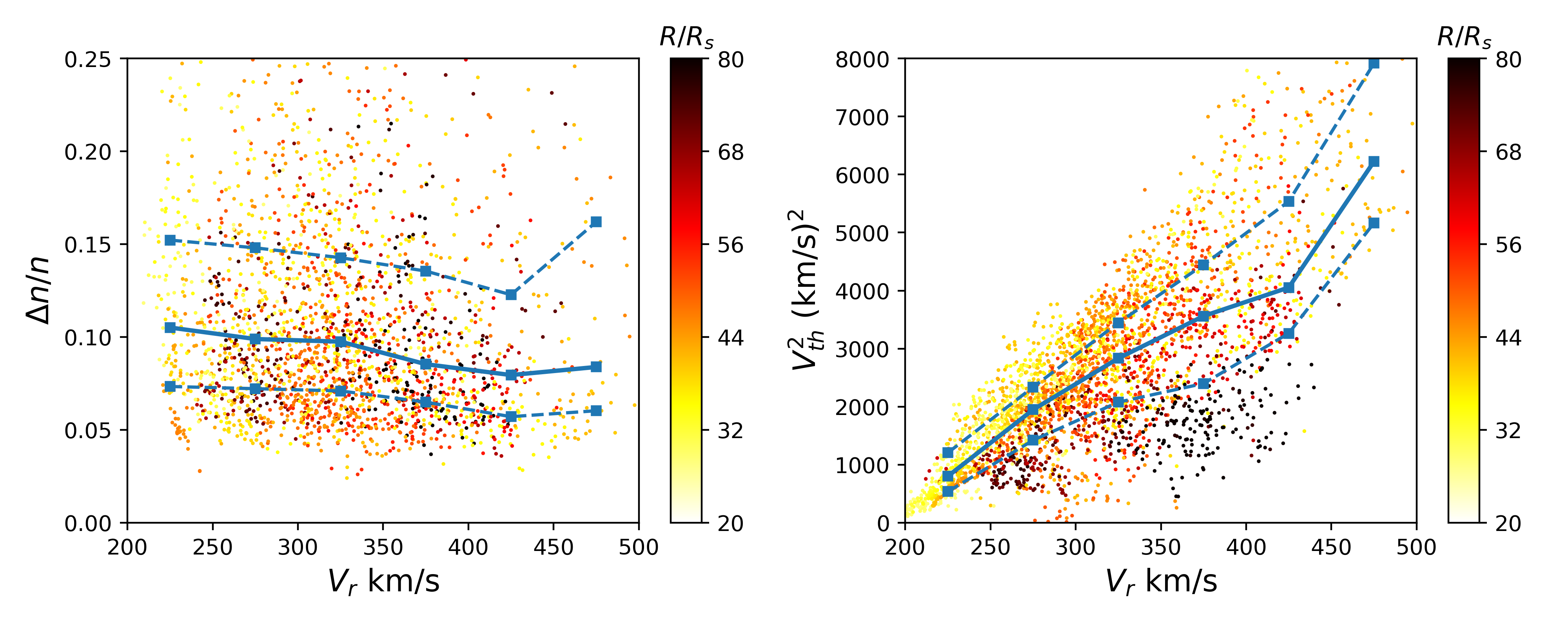}
    \caption{Relative density fluctuation $\Delta n / n$ (left) and ion temperature (right) expressed in thermal speed squared as functions of radial solar wind speed. Each dot corresponds to a single half-hour window and the colors represent the radial distance to the Sun. Squares on solid curves are median values of the dots binned according to $V_r$, and squares on dashed curves are the other two quartiles.}
    \label{fig:statistic_n_fluc_T_vr}
\end{figure*}

There are several points that are worth underscoring here. (1) From Row (g)\&(h), we can see that both magnetic and kinetic energies in the waves decrease as we move away from the Sun. This is a natural result, mainly of the spherical expansion of the solar wind but also of the energy cascade of the turbulence. A similar trend of $E_{o}$ and $E_i$ was reported in \citep{chen2020evolution}. (2) The streams measured by PSP during Encounters 4\&5 are mostly of a very low speed (Row (b)). As is shown in Section \ref{sec:acceleration_SW}, the streams are actually still accelerating radially. (3) The ion thermal speed (Row (b)), or equivalently the square-root of ion temperature, is highly correlated with the radial flow speed, which is a well-known phenomenon that has already been observed by other satellites \citep{Grappinetal1990,demoulin2009temperature} and the PSP measurements show that this correlation exists at radial distances even down to $\sim 28$ solar radii. A statistical analysis of this point is presented in Section \ref{sec:density_fluc_T_Vr}. (4) The density profile (Row (c)), except for a slow variation with the radial distance, shows strong structures near the perihelion during Encounters 4\&5. For example, a short plasma sheet crossing was observed on January 30, 2020 and a long plasma sheet crossing was observed on June 8, 2020. These structures have significant impacts on the turbulence properties as is discussed in detail in Section \ref{sec:discussion}. (5) The slopes of the magnetic field and velocity spectra (Row (d)) highly fluctuate and show a dependence on the stream properties. It can be observed from Fig. \ref{fig:overview} that the magnetic field spectrum is usually steeper than the velocity spectrum, especially far from the Sun. Near perihelion, the two slopes seem to be close to each other. A statistical survey of spectral variability is presented in Section \ref{sec:evolution_spectra}. (6) Usually, the normalized cross helicity $\sigma_c$ (Row (e)) is close to 1, implying a status of dominating outward-propagating Alfv\'en waves. However, there are periods when $\sigma_c$ oscillates and becomes negative, for example, November 10, 2018, January 17-21, 2020, and June 8, 2020. As is shown in Section \ref{sec:discussion}, these periods correspond to PSP observing heliospheric large-scale inhomogeneous structures, such as velocity shears and the heliospheric current sheet. Furthermore, $\sigma_c$ is also found to be significantly smaller throughout E5 when compared to E1\& E4 and the reason for this is also discussed in Section \ref{sec:discussion}. (7) We note that  $\sigma_r$ (Row (f)) shows interesting behavior: During Encounter 1, its value is very close to zero, indicating balanced magnetic and kinetic energies, which are expected for Alfv\'enic turbulence. However, during Encounters 4\&5, most of the time it is negative, especially close to perihelion. This suggests the possibility that the turbulence is magnetically-dominated at its origin inside certain types of streams. Statistical analyses are presented later in Section \ref{sec:evolution_spectra}.

\subsection{Evidence of accelerating solar wind streams}\label{sec:acceleration_SW}
\begin{figure*}
    \centering
    \includegraphics[width=\textwidth]{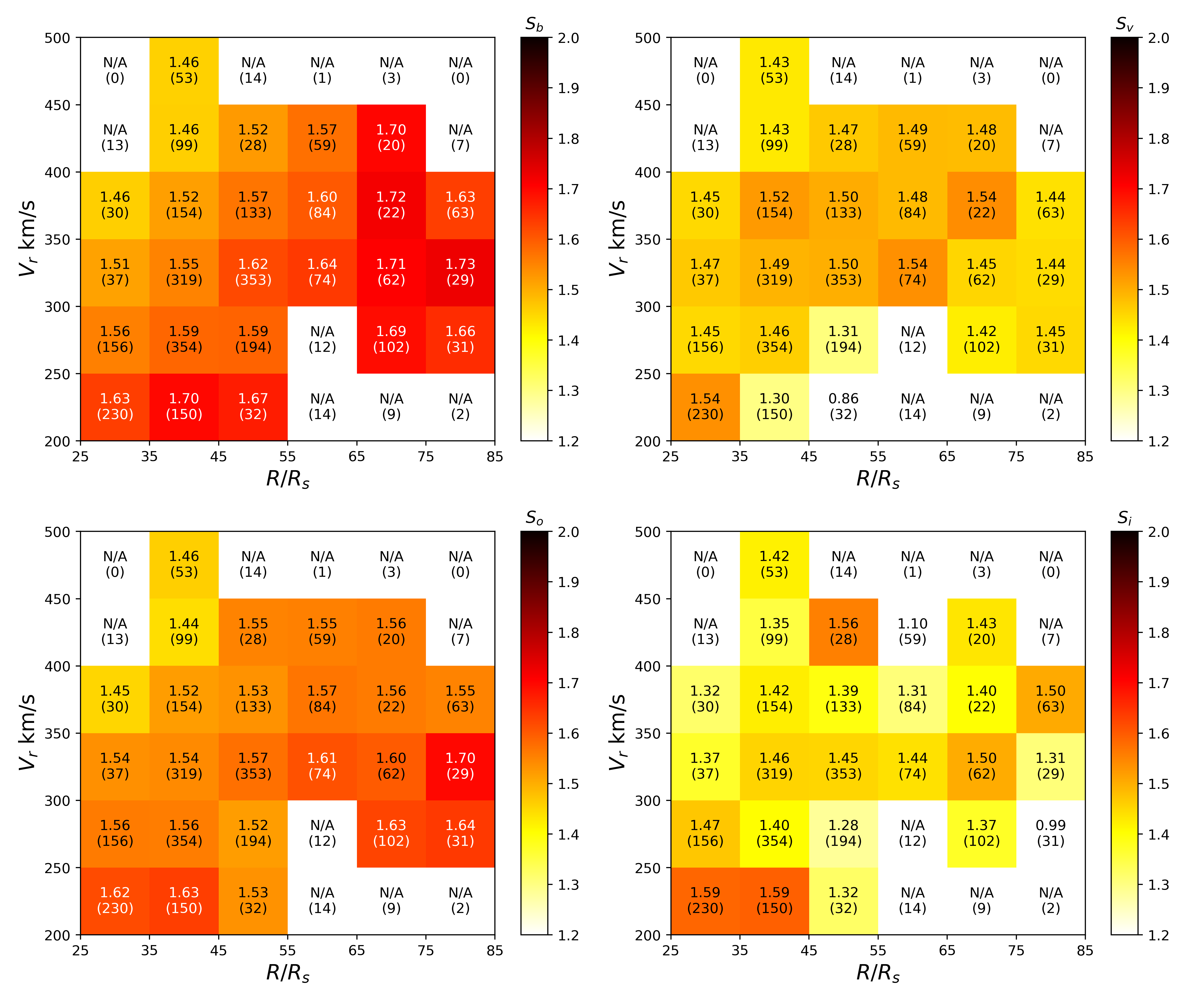}
    \caption{Spectral slopes of the magnetic field in Alfv\'en unit $\mathbf{B}/\sqrt{\mu_0 \rho}$ (top left), velocity (top right), outward Els\"asser variable (bottom left), and inward Els\"asser variable (bottom right) as functions of the radial solar wind speed $V_r$ and radial distance to the Sun $R$. The data points were binned according to $V_r$ and $R$, and the median value inside each bin was calculated, which is reflected in the colors and written in the plot. The bracketed numbers in the plots are the number of data points inside each bin. Bins with no more than 15 data points were discarded. }
    \label{fig:spectral_slopes_2D}
\end{figure*}

\begin{figure*}
    \centering
    \includegraphics[width=\textwidth]{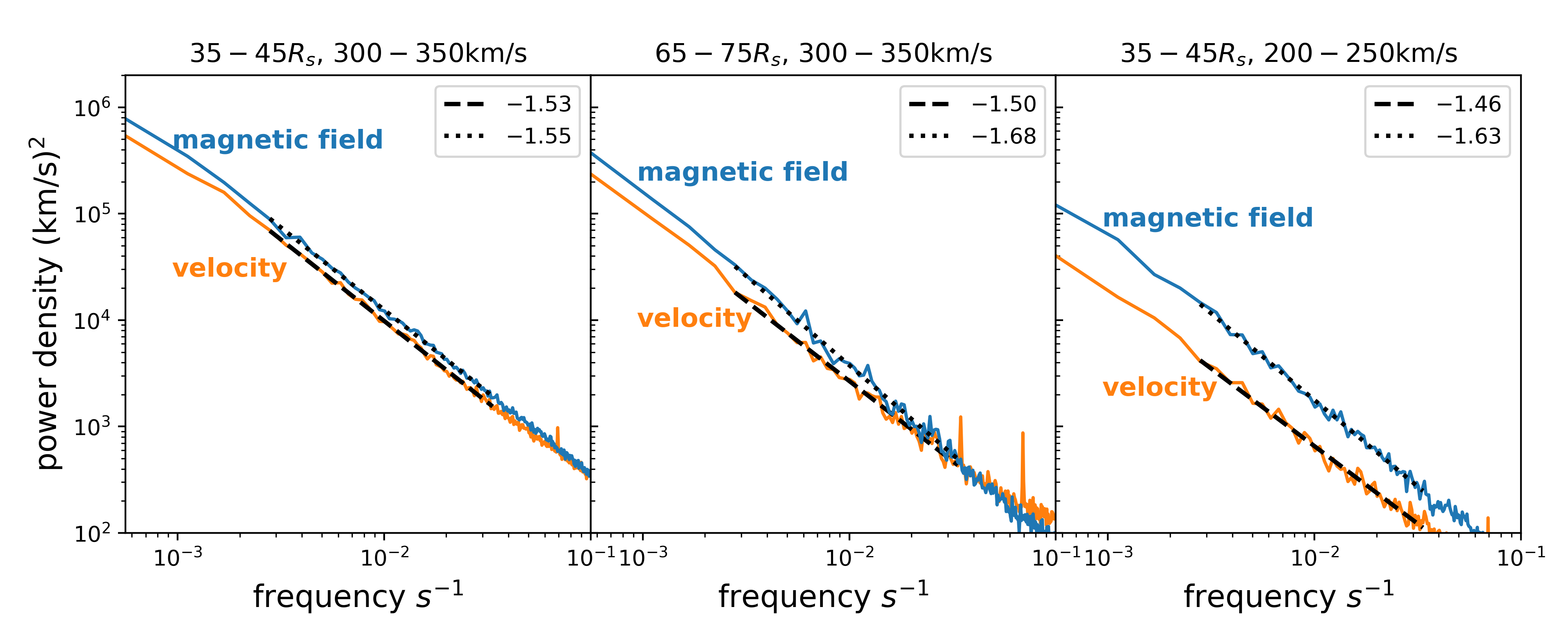}
    \caption{Averaged power spectra of magnetic field (in Alfv\'en speed) and velocity for different $R$ and $V_r$. Left: $35 \leq R/R_s \leq 45$ and 300km/s$\leq V_r \leq$350km/s. Middle: $65 \leq R/R_s \leq 75$ and 300km/s$\leq V_r \leq$350km/s. Right: $35 \leq R/R_s \leq 45$ and 200km/s$\leq V_r \leq$250km/s. The spectra were fitted over $2.8\times 10^{-3} s^{-1} \leq f \leq 1.7 \times 10^{-2} s^{-1}$ as shown by the dotted (for magnetic field) and dashed (for velocity) lines and the fitted slopes are written in the legend.}
    \label{fig:average_spectrum}
\end{figure*}

\begin{figure*}
    \centering
    \includegraphics[width=\textwidth]{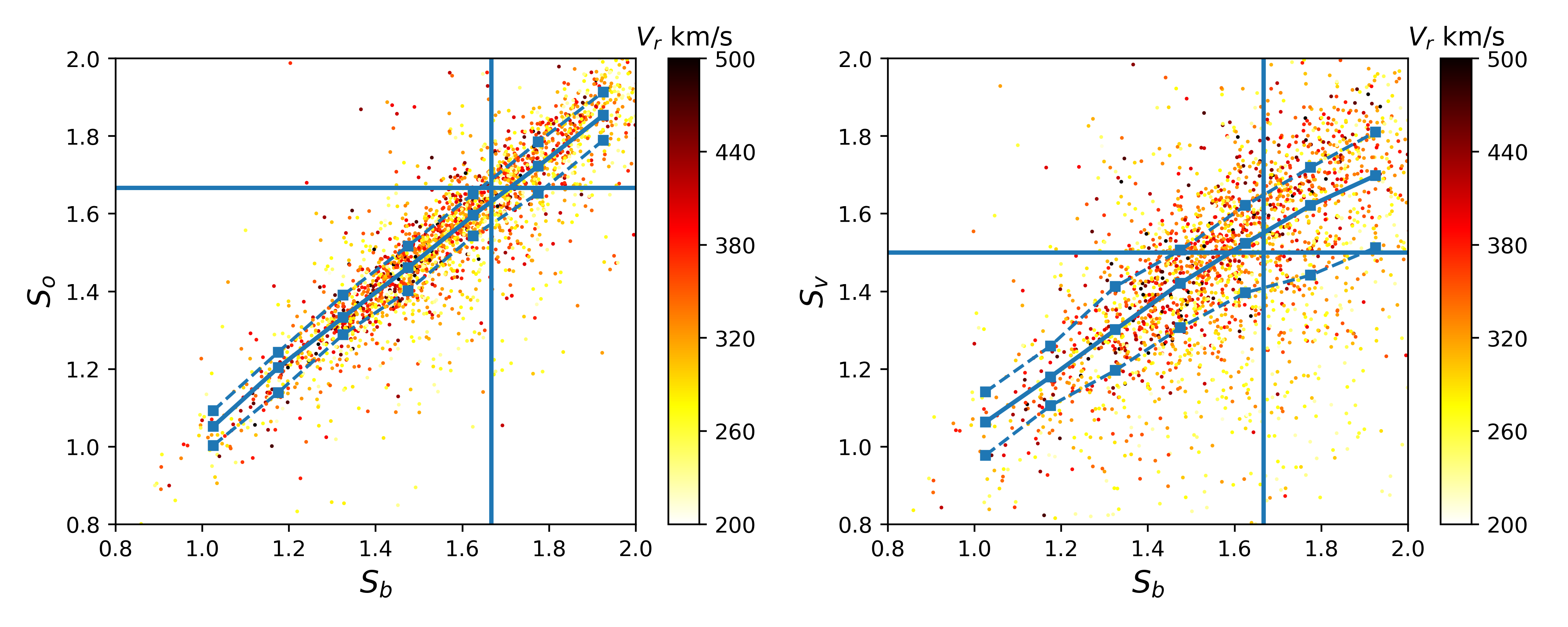}
    \caption{Left: Spectral slope of outward Els\"asser variable $S_o$ as a function of the spectral slope of magnetic field $S_b$. Right: Spectral slope of velocity $S_v$ as a function of the spectral slope of magnetic field $S_b$. Vertical lines mark $S_b = 5/3$. The horizontal line in the left panel marks $S_o = 5/3$ and the horizontal line in the right panel mark $S_v = 3/2$. The colors represent the radial speed of the solar wind.}
    \label{fig:correlation_slopes}
\end{figure*}

As PSP travels to a sufficiently low altitude above the Sun, its relative longitudinal speed to the rotating solar surface changes sign when it crosses a critical height. That is to say, in the reference frame corotating with the Sun, PSP moves toward the west first as its altitude lowers, then it retrogrades to the east near the perihelion and finally moves back toward the west as it goes away from the perihelion. This unique feature of PSP's orbit makes it possible to conduct a better analysis of the spatial structures in the solar wind as PSP may measure streams from the same region on the solar surface for two or three times at different radial distances to the Sun during one encounter. 

In Fig. \ref{fig:3D_trajectory_E04}, we show the measurements of the radial magnetic field, radial flow speed, and proton number density during Encounter 4 in the top three panels. Instead of plotting these quantities against time, we plotted them on a radius-(Carrington longitude) grid so that the projection of the curves on the grid is approximately the trajectory of PSP in the reference frame corotating with the Sun. We note that the inclination of PSP's orbit is very low: The Carrington latitude of PSP during Encounter 4 varies between $\sim \pm 4^\circ$. In the bottom panel, we plotted the trajectory of PSP for reference purposes with $z$-axis being the Carrington latitude. The colors of the curves represent the time such that PSP travels from the light-colored end toward the dark-colored end. From the $V_r$ plot, we can clearly see that the stream measured near perihelion (the dark branch of the curve) contains similar structures observed in the stream further away from the Sun (the light branch of the curve). This similarity can also be seen in the $B_r$ and $n_p$ plots, implying that the satellite observed streams coming from the same region of the Sun twice as it traveled inward and outward during the encounter. In the top panel of Fig. \ref{fig:Vr_carr_long_E04}, we show a 2D $V_r$-longitude plot so that we can better compare the measurements made as PSP traveled back-and-forth in longitude. Despite some deformation, the two curves show very similar variations. We marked the identified similar structures at different radial distances by the red circles connected by dashed lines and the radial wind speed at these circles is plotted against the radial distance in the bottom panel, where each line corresponds to one single structure, which is numbered as 1, 2, 3, 4, 5, and 6 as annotated in the top panel. We can see the wind is accelerated at a rate of $1-3$ km/s/$R_s$ for most of these structures and overall the streams are accelerated from $\sim (200-300)$ km/s near perihelion ($\sim 30 R_s$) to $\sim (300-400)$ km/s beyond $60 R_s$. Thus, these kinds of measurements made uniquely by PSP can be used to quantify the acceleration of the solar wind in the future with increasing data volume. A couple of caveats should be noted here. First, in the corotating frame of the Sun, there is an additional longitudinal speed of the wind. Thus, the stream measured by PSP at a certain longitude should come from a region located at a larger longitude on the solar surface. For example, a 300 km/s wind drifts along longitude by $\sim 17^\circ$ after it propagates $50R_s$. Apparently, streams of different velocities and measured at different distances should drift by different amounts in longitude. In addition, considering the wind is accelerating, it is even more complicated to estimate this longitudinal drift. Second, the variation of the latitude of PSP, though not very large, may also account for the deformation of the longitudinal profiles of the measured streams.

\begin{figure*}
    \centering
    \includegraphics[width=\textwidth]{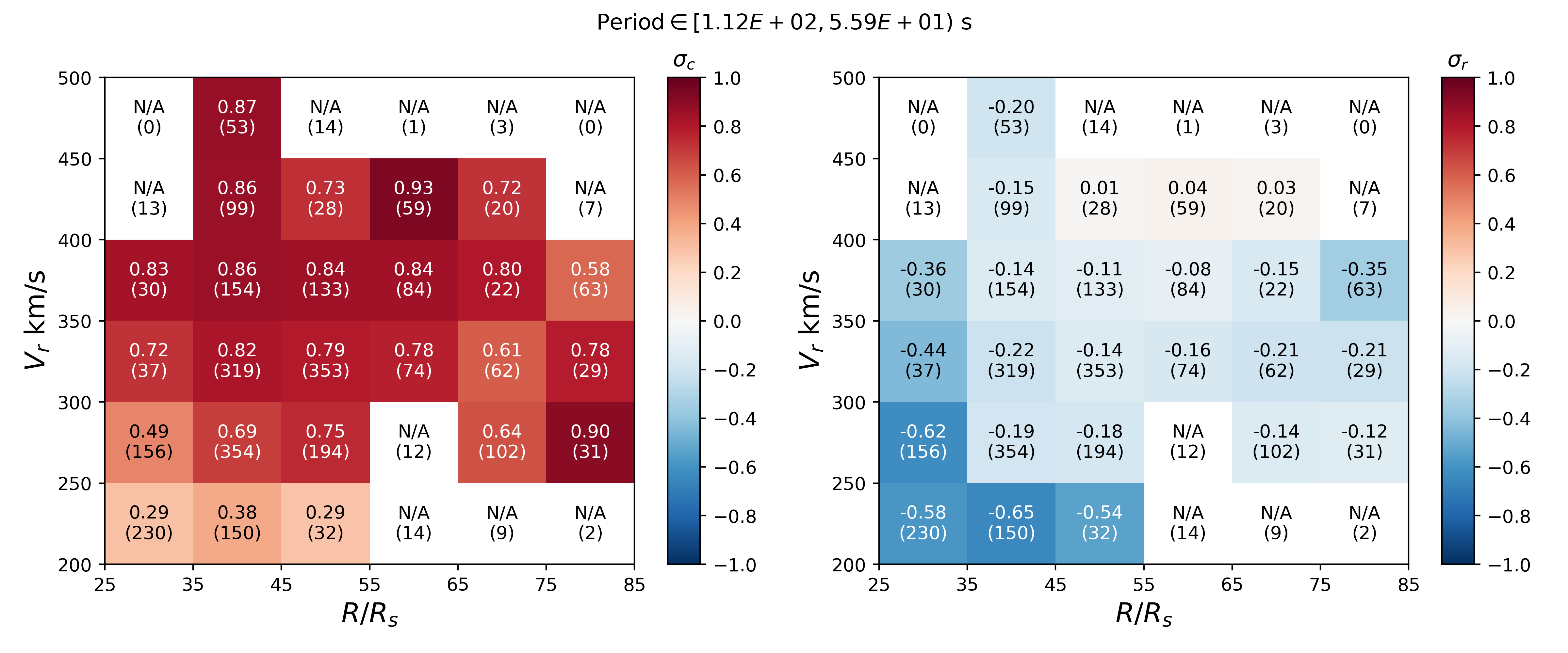}
    \caption{Normalized cross helicity $\sigma_c$ (left) and normalized residual energy $\sigma_r$ (right) of wave band 5 ($T \approx 112-56 $ s) as functions of the radial distance to the Sun $R$ and the radial speed of solar wind $V_r$. The colors of each block represent the median values of the binned data. Text on each block shows the value of the block and the number of data points (in brackets) in the block. Bins with no more than 15 data points were discarded.}
    \label{fig:sigma_c_sigma_r}
\end{figure*}

\subsection{Dependence of density fluctuations 
and ion temperature on the wind speed}\label{sec:density_fluc_T_Vr}
In Fig. \ref{fig:statistic_n_fluc_T_vr}, we show the scatter plot of the relative density fluctuation $\Delta n / n$ and temperature $V_{th}^2$ versus the radial flow speed $V_r$. Each single dot represents a half-hour time window and the color of each dot shows the radial distance to the Sun. We note that $V_{r}$, $V_{th}$, and $n$ are averaged over the time window while $\Delta n $ is the standard deviation of $n$ inside the window. To better show the trend, we binned the dots with 50 km/s $V_r$ intervals and calculated the three quartiles inside each bin, which are shown as the blue squares. 

\begin{figure*}
    \centering
    \includegraphics[width=\textwidth]{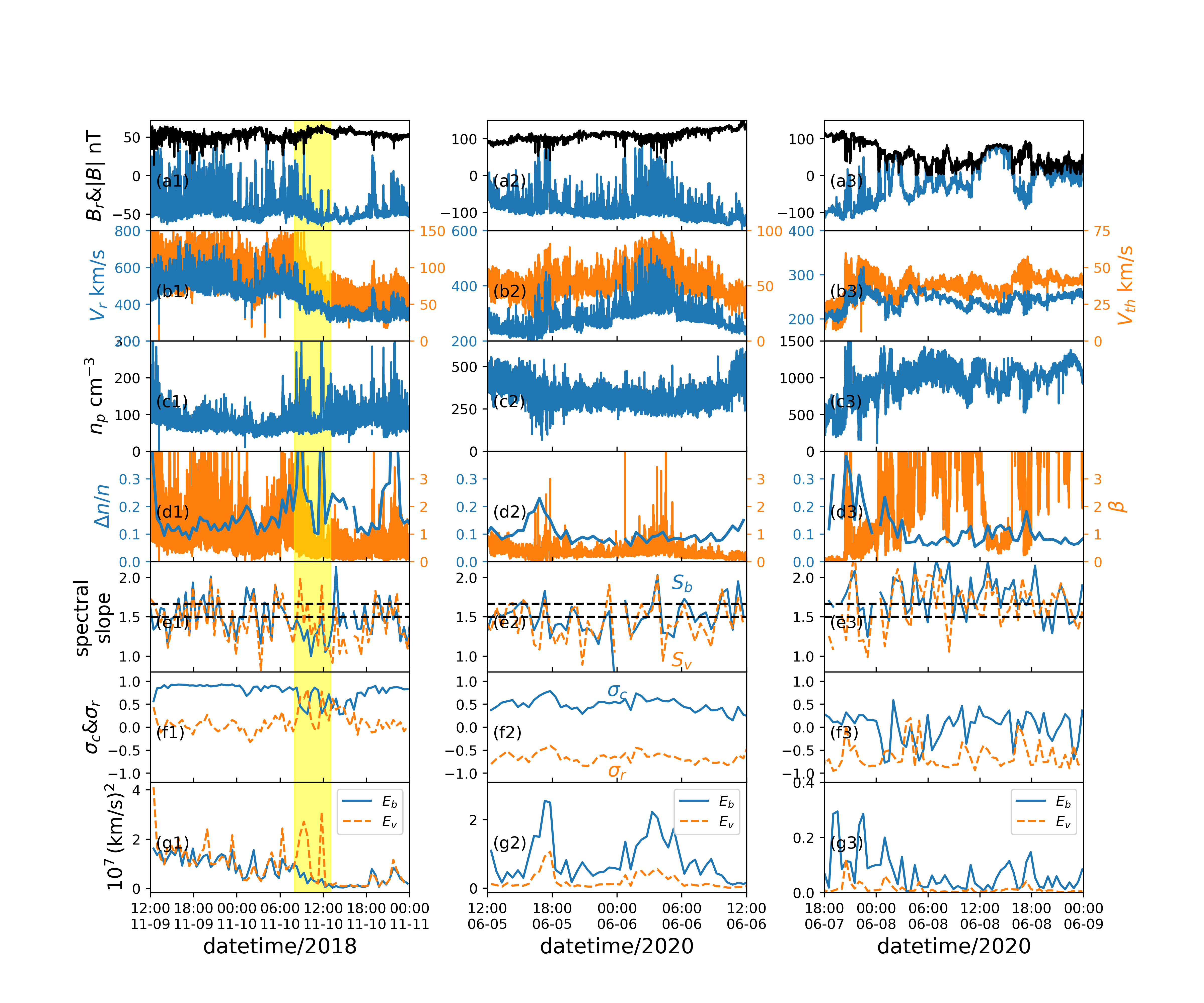}
    \caption{Blow-ups of the time periods marked by the shaded regions in Fig. \ref{fig:overview}. Row (a) shows the radial magnetic field $B_r$ (blue) and the magnitude of magnetic field $|B|$ (black). Row (b) shows the radial flow speed $V_r$ (blue) and the ion thermal speed $V_{th}$ (orange). Row (c) shows the ion density $n_p$. Row (d) shows the relative ion density fluctuation $\Delta n/n$ (blue) and the plasma beta $\beta$ (orange), defined as the ion thermal pressure $p_{th}=n_p m_i V_{th}^2$ divided by magnetic pressure $p_{mag}=B^2/2\mu_0$. Row (e) shows the spectral slopes of magnetic field in Alfv\'en speed (blue) and velocity (orange). The two dashed lines mark $3/2$ and $5/3$ for reference. Row (f) shows $\sigma_c$ (blue) and $\sigma_r$ (orange). Row (g) shows the energies in magnetic field fluctuations $E_b$ (blue) and velocity fluctuations $E_v$ (orange).}
    \label{fig:blow_up}
\end{figure*}
\begin{figure}
    \centering
    \includegraphics[width=\hsize]{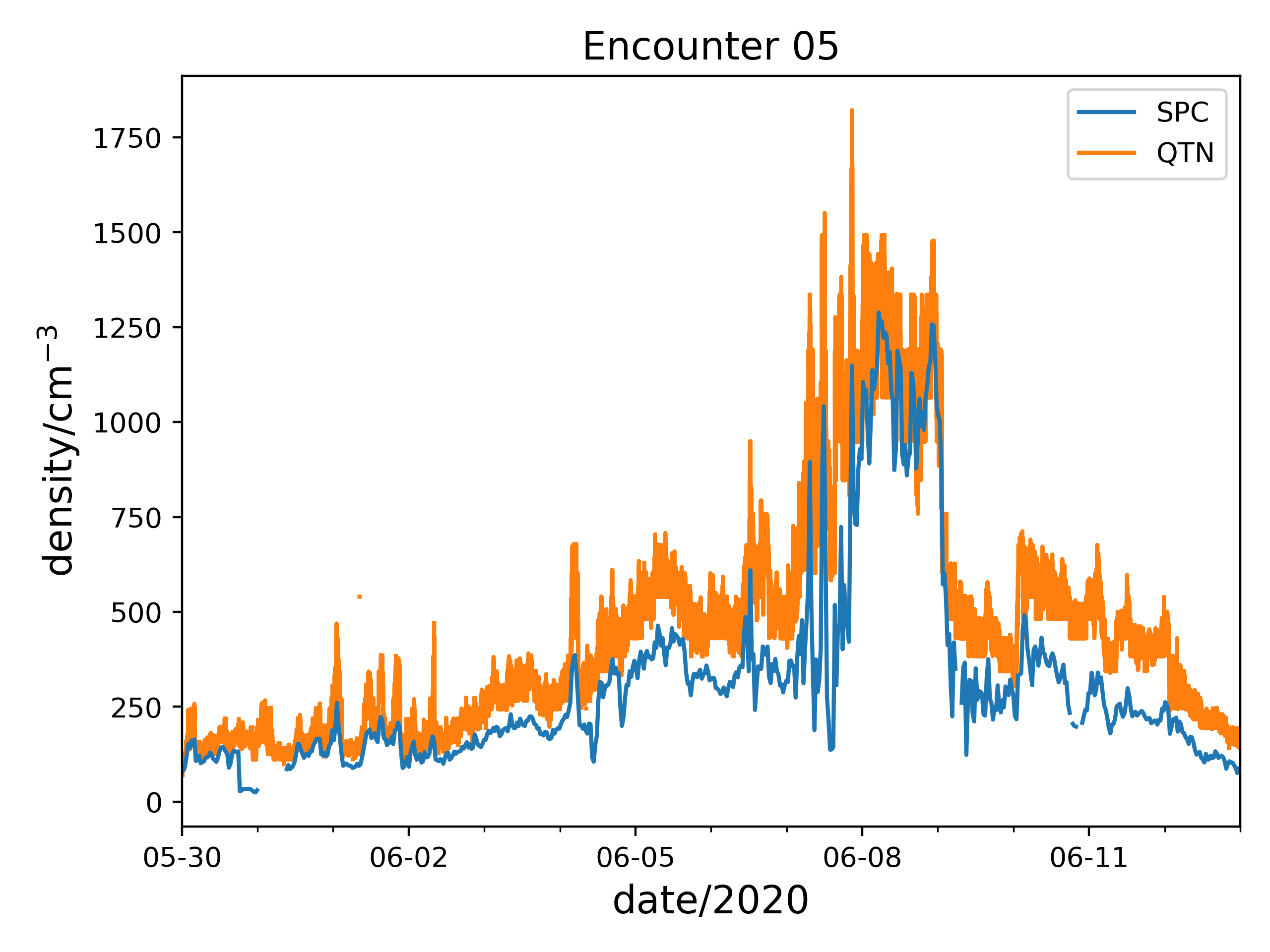}
    \caption{Comparison between the ion and electron densities during Encounter 5. Blue: Ion density measured by the Faraday cup (SPC). Orange: Electron density calculated using the quasi thermal noise (QTN) measurements made by the Radio Frequency Spectrometer Low Frequency Receiver (RFS/LFR).}
    \label{fig:compare_SPC_QTN}
\end{figure}
\begin{figure*}
    \centering
    \includegraphics[width=\textwidth]{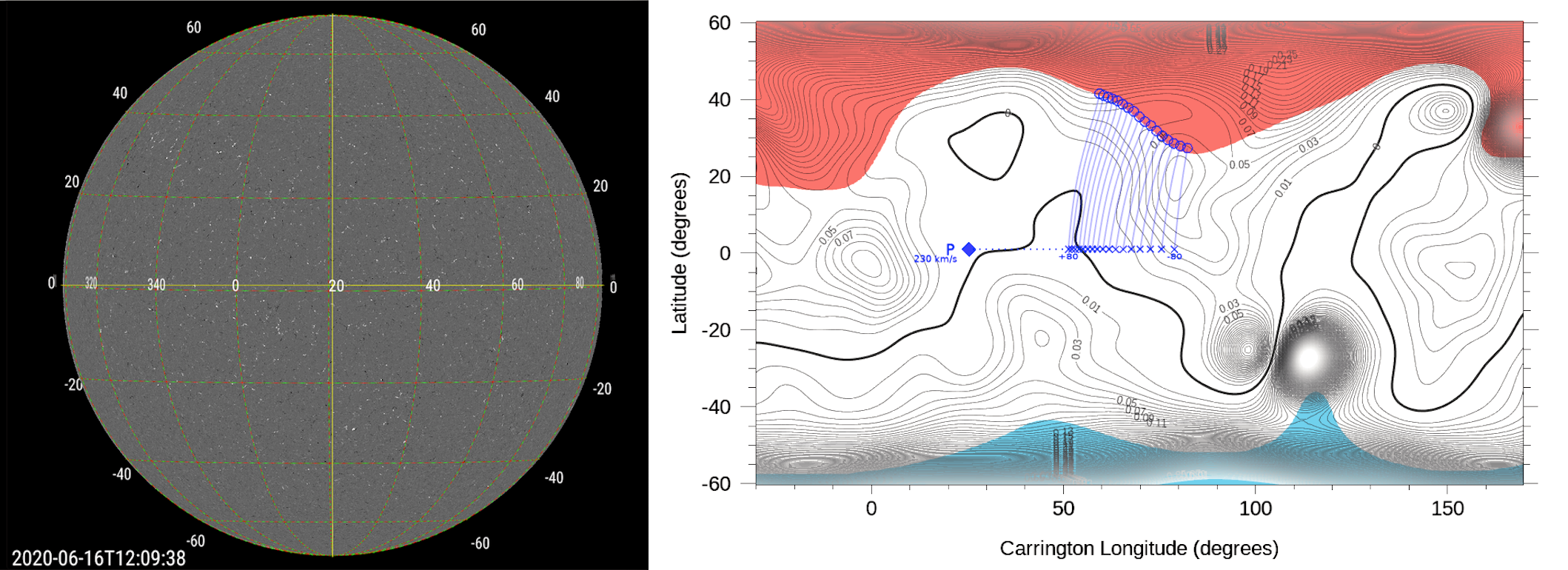}
    \caption{Left: SDO/HMI image taken on  Jun 16, 2020, corresponding to Encounter 5 of PSP. The grid is in Carrington degrees. One can see that during Encounter 5, the visible side of the Sun was very quiet. Right: Magnetic pressure map at $R=1.2R_s$ calculated by the PFSS model with the source surface at $R_{ss}=2.5R_s$ and SDO/HMI data on Jun 16, 2020 as input. The blue diamond is the direct radial projection of PSP to the source surface. The blue crosses are the foot points of the magnetic field lines connected to PSP on the source surface. Different blue crosses correspond to a prediction using varying wind speeds, from $230- 80$km/s to $230+80$km/s. The blue circles are on the surface $R=1.2R_s$ and are magnetically connected to the blue crosses. The thick black lines are the neutral lines at $R=1.2R_s$, and colored regions are the open magnetic field regions with blue being negative polarity and red being positive polarity.}
    \label{fig:connection_to_corona}
\end{figure*}

Although the value of $\Delta n/n$ is pretty scattered, it is in general small, mostly smaller than 0.2, and it decreases with $V_r$. We note that the rise of the blue squares at $V_r \in [450,500]$ km/s is very likely a result of a lack of data points. From the colors of the dots, we cannot see a clear relation between $\Delta n/n$ and $R$. Thus, we conclude that the density fluctuation is larger in the slow streams than the fast streams and it does not evolve significantly as the solar wind propagates. The ion temperature is less scattered than $\Delta n / n$ and the $V_{th}^2 - V_r$ relation shows very good linearity, as already mentioned in Section \ref{sec:overview}. This strong $T-V$ correlation is a well-known phenomenon observed at 1 AU \citep[e.g.,][]{elliott2005improved,matthaeus2006correlation,demoulin2009temperature} and PSP data show that this correlation is already well established as close as 30 $R_s$. This may be a clue as to the origin of this $T-V$ correlation. \citet{matthaeus2006correlation} proposed that this correlation is a result of the fact that the transport equation of temperature with a constant radial speed $V$ has a solution of the form $T=T(R/V)$. Thus, in supposing $T$ is a decreasing function, we expect that a larger radial speed leads to a slower decay of $T$ with $R$, resulting in the observed positive $T-V$ correlation. On the other hand, \citet{demoulin2009temperature} argued that this correlation is a requirement by the momentum equation as a higher temperature is needed to accelerate the solar wind to a higher speed. Since the measurements made during the encounters of PSP are likely in the accelerating solar wind streams as pointed out in Section \ref{sec:acceleration_SW}, it is reasonable to say that the origin of the positive $T-V$ correlation is related to the acceleration mechanism of the solar wind. A better modeling of the solar wind heating and acceleration is necessary to fully understand this issue. Last, from the right panel of Fig. \ref{fig:statistic_n_fluc_T_vr}, it seems that very close to the Sun (light yellow dots), the slope of the $T-V$ relation is larger than that further away from the Sun (dark red dots). If it is true that the $T-V$ slope changes radially, it implies that the adiabatic cooling rate is a function of the solar wind speed, which is true for electrons \citep{maksimovic2020anticorrelation}. However, we should be cautious in making this conclusion because during different encounters, the solar condition might be very different.


\subsection{Evolution of the turbulence spectra and Alfv\'enicity}\label{sec:evolution_spectra}
As already described in Section \ref{sec:inst_data}, we calculated the spectral slopes over a period range $T \in [30s, 360s]$ for the magnetic field in Alfv\'en speed unit, velocity, and outward and inward Els\"asser variables. The statistical results of these slopes are presented in Fig. \ref{fig:spectral_slopes_2D}. We binned the data points according to the radial solar wind speed $V_r$ and the radial distance to the Sun and then calculated the median value inside each bin. The median values are reflected by the colors of the blocks in Fig. \ref{fig:spectral_slopes_2D} and are also written in the blocks. The bracketed numbers in the plots are the number of data points and we discarded the bins with no more than 15 data points (values were set to N/A).

We first compared the top two panels of Fig. \ref{fig:spectral_slopes_2D}, that is to say the spectral slopes of the magnetic field ($S_b$) and velocity ($S_v$). There is no clear $V_r$-dependence of $S_v$, while a negative $S_b$-$V_r$ correlation is observed in the range of $R \in [35, 65]R_s$. Close to the Sun ($R<45R_s$), the difference between $S_b$ and $S_v$ is small. Both of the magnetic field and velocity spectra are flatter than the Kolmogorov’s prediction -5/3 and are around -1.5. As the radial distance increases, steepening of magnetic field spectrum toward a -5/3 slope is seen while the velocity spectrum slope remains quite constant. In Fig. \ref{fig:average_spectrum}, we plotted the power spectra of the magnetic field (in Alfv\'en speed) in blue and the velocity in orange, averaged over all half-hour windows that fall into a specific radial distance range and wind speed range. We fit the spectra over the period range $T\in[30,360]s$ and the fitted slopes are written in the plot. The left panel is for $R\in[35,45]R_s$ and $V_r \in[300,350]$km/s and the two spectra have nearly identical slopes close to $-1.5$. The middle panel is for $R\in[65,75]R_s$ and $V_r \in[300,350]$km/s and it shows that as $R$ increases to around 0.3 AU, the magnetic field spectrum becomes close to the Kolmogorov's spectrum while the velocity spectrum is still the Iroshnikov-Kraichnan spectrum. The right panel is for $R\in[35,45]R_s$ and $V_r \in[200,250]$km/s and by comparing it with the left panel, we can see that at the same radial distance $R$, the slower wind has a steeper magnetic field spectrum. It has been long observed outside 0.3 AU that the magnetic field spectrum is steeper than the velocity spectrum \citep[e.g.,][]{Grappinetal1991}. Figures \ref{fig:spectral_slopes_2D}\&\ref{fig:average_spectrum} suggest that very close to the Sun, the two spectra may have the same slope. The anticorrelation between $S_b$ and $V_r$ and the positive correlation between $S_b$ and $R$ imply the existence of a ``turbulence age'' which determines the level of the turbulence development. A recent work analyzing Helios, Wind, and Ulysses data reveals similar ``aging'' of turbulence radially beyond 0.3 AU \citep{weygand2019jensen}. From the bottom two panels of Fig. \ref{fig:spectral_slopes_2D}, the spectral slope of $\mathbf{Z_{o}}$ shows a similar evolution with that of the magnetic field, but it is shallower. The spectral slope of $\mathbf{Z_i}$, on the other hand, resembles the velocity, that is to say it does not show significant radial evolution and it is even smaller than the velocity slope. In Fig. \ref{fig:correlation_slopes} we show the correlation between $S_o$ and $S_b$ in the left panel and the correlation between $S_v$ and $S_b$ in the right panel. Both of the two correlations are high, especially that between $S_o$ and $S_b$. The $S_v-S_b$ correlation is weaker than the $S_o - S_b$ correlation due to the fact that $S_b$ varies with $V_r$ and $R$, while $S_v$ is quite constant. For reference purposes, we marked $S_b = 5/3$ by the vertical lines and $S_o = 5/3$, $S_v = 3/2$ by the two horizontal lines. We see that, on average, $S_o$ is close to $S_b$, though slightly smaller, while $S_v$ is clearly smaller than $S_b$ such that $S_b = 5/3$ corresponds to an $S_v$ around 1.55-1.6. This result is similar to that reported by \citet{Grappinetal1991} (see their Figure 7), although the data used here are mainly
within 0.3 AU, while \citet{Grappinetal1991} analyzed Helios data that were collected outside 0.3 AU.

Figures \ref{fig:spectral_slopes_2D}\&\ref{fig:average_spectrum} reveal that in the very young solar wind, the magnetic field and velocity spectra have the same slope; furthermore, as the turbulence evolves, the magnetic field spectrum steepens while the velocity spectrum has an invariant slope. This poses a challenge in understanding the nature of the MHD turbulence in the solar wind. Most of the turbulence theories \citep[e.g.,][]{Kraichnan1965,Goldreich1995,lithwick2003imbalanced,Zanketal2017} describe the turbulence based on the two Els\"asser variables, thus they cannot directly capture the differential evolution of the magnetic field and velocity spectra. \citet{boldyrev2011spectral} conducted 3D incompressible MHD simulations based on the reduced equation set of Els\"asser variables and they reproduced the different magnetic field and velocity spectra statistically. However, how the final status is established is still unknown from the simulations. In addition, PSP data show that the steepening of the magnetic field spectrum is quite slow. The top-left panel of Fig. \ref{fig:spectral_slopes_2D} implies the steepening from 3/2 to 5/3 takes time for the wind to travel from $R\sim 30R_s$ to $R \sim 70 R_s$. This is much longer than the nonlinear time, that is the ``eddy-turnover'' time or the Alfv\'en crossing time, of the turbulence. Thus, it is possible that in the solar wind, the differential evolution of $B$ and $V$ is controlled by some external mechanisms, such as stream shears and the spherical-expansion effect, which leads to different decay rates of the magnetic energy and kinetic energy \citep{GrappinandVelli1996}.

In Fig. \ref{fig:sigma_c_sigma_r}, we present the $(V_r,R)$ variation of the normalized cross helicity $\sigma_c$ (left panel) and the normalized residual energy $\sigma_r$ (right panel), in a similar manner as we do for Fig. \ref{fig:spectral_slopes_2D}. Here the values were calculated for the wave band 5, that is corresponding to wave period $T\in[112,56]$s, while other wave bands show similar features as shown in Fig. \ref{fig:sigma_c_sigma_r}. For $\sigma_c$, an overall positive $\sigma_c$-$V_r$ correlation is observed, at least for $R \leq 65 R_s$, indicating that the fast wind is generally more Alfv\'enic than the slow wind. The lack of a definite $\sigma_c-V_r$ correlation for $R>65  R_s$ might be due to the lack of data points so that the value in one single block mainly reflects the turbulence property inside one stream instead of multiple streams, increasing the uncertainty. The $\sigma_c-R$ correlation is clearly negative in the range of $R \ge 35R_s $ and $V_r \in [300,400]$km/s, implying that the dominance of the outward propagating wave declines with the radial distance, which was already reported in previous works \citep[e.g.,][]{chen2020evolution}. But this correlation is not well-defined in other parametric regions. Especially, for measurements made below $35R_s$ and for very slow wind ($V_r \leq 250 $km/s), $\sigma_c$ is much lower compared with the neighboring blocks in $V_r-R$ space. This is caused by the non-Alfv\'enic, or low-Alfv\'enic, slow wind measured by PSP during Encounter 5 (see right column of Fig. \ref{fig:overview}). For $\sigma_r$, we can see that it is in general negative, that is to say the magnetic energy exceeds the kinetic energy, which is a well-known phenomenon that is not fully understood yet. For $R \leq 65 R_s$, $\sigma_r$ is also positively correlated with $V_r$. That is to say, in the fast wind, the magnetic and kinetic energies are more balanced, which is consistent with the high $\sigma_c$ values which imply a highly Alfv\'enic status. The radial evolution of $\sigma_r$, however, shows a surprising result as it is clear that inside $65R_s$, $\sigma_r$ increases with radial distance, meaning that the turbulence is relaxing from a magnetic-dominating status toward a more balanced status. Actually, by examining the middle and right columns of Fig. \ref{fig:overview}, one can find that $\sigma_r$ is clearly an increasing function of $R$ from January 21-29, 2020 and from June 1-6, 2020, which is consistent with the statistical result here. Even for Encounter 1 (left column of Fig. \ref{fig:overview}), a slight increase in $\sigma_r$ with $R$ is observed from November 6-9, 2018. Outside $65R_s$, the evolution is not very clear but it seems that $\sigma_r$ may start to drop with $R$. Similar to $\sigma_c$, the values of $\sigma_r$ are extremely low for $R\leq 35R_s$ and for $V_r \leq 250 $km/s. As mentioned before, this region in the parameter space corresponds to the very low Alfv\'enic streams observed during Encounter 5.

\section{Discussion}\label{sec:discussion}
From Section \ref{sec:evolution_spectra}, we conclude the following points: (1) During the evolution of the solar wind turbulence, the magnetic field spectrum steepens from a $-3/2$ slope toward a $-5/3$ slope while the velocity spectrum slope remains $-3/2$. (2) The fast solar wind is in general more Alfv\'enic than the slow solar wind, with $\sigma_c$ closer to 1 and $\sigma_r$ closer to 0. However, we should emphasize here that the ``fast'' solar wind in this study is not the typical fast wind that originates from large-scale polar coronal hole open regions because during the first five encounters, PSP did not observe any long-lasting fast solar wind of this type. Thus, it is more likely that the ``fast'' winds here should probably be classified as examples of ``faster'' Alfv\'enic slow wind \citep{DAmicisandBruno2015,panasenco2020exploring}. (3) Closer to the Sun, $\sigma_c$ increases toward 1, confirming that the turbulence is dominated by outward propagating Alfv\'en waves in the young solar wind. However, there are periods where $\sigma_c$ is quite low even at very close distances to the Sun (below $35R_s$). (4) For some solar wind streams, for example, those observed during Encounters 4\&5, $\sigma_r$ evolves from negative values toward 0 at close distances, suggesting that the turbulence is actually magnetic-dominated at its origin and then gradually relaxes to a more balanced status in these streams.

The above conclusions are based on the statistical results using all high-resolution data from PSP's first five encounters. While they help us depict an average picture of the evolution of solar wind turbulence, it is still necessary to examine the turbulence from different time periods so that we can have deeper insights on how the turbulence varies in different streams. In fact, as one may have noticed in Fig. \ref{fig:overview}, fluctuations in streams of a similar radial wind speed can have significantly different Alfv\'enicity. For example, during E1 from November 3-7, the solar wind speed is around 300km/s and the fluctuations are highly Alfv\'enic, while during E5 from June 1-6, the solar wind speed is also around 300km/s but the Alfv\'enicity of the fluctuations is quite low. A more detailed analysis is presented later in this section. In Fig. \ref{fig:blow_up}, we present the blow-ups of Fig. \ref{fig:overview} over three short time periods marked by the shades in Fig. \ref{fig:overview}. Compared with Fig. \ref{fig:overview}, the top three rows of Fig. \ref{fig:blow_up} present data at a time resolution of 0.874s instead of a half hour. In addition, in the bottom two rows of Fig. \ref{fig:blow_up}, $\sigma_c$, $\sigma_r$, $E_{b}$, and $E_v$ were calculated by integrating over all wave modes except mode 0, that is to say the background field.

\subsection{Alfv\'enic turbulence and the effect of velocity shear}
The left column of Fig. \ref{fig:blow_up} shows the time period from 12:00 November 9 to 00:00 November 11, 2018 during Encounter 1. Before 08:00 November 10, PSP was inside a fast stream with a radial speed of $V_r \sim 500-600$km/s. Between 08:00 and 13:00 November 10, PSP crossed a fast-slow stream shear region, marked by the shaded region, after which the wind speed dropped to less than 400km/s. Inside the fast stream, a large amount of switchbacks were observed with nearly constant $|B|$ and $n_p$, as well as $\sigma_c \approx 1$ and $\sigma_r \approx 0$. These parameters imply that the turbulence is highly Alfv\'enic, with very little inward propagating wave component. Inside the shear region, a decrease in $\sigma_c$ and increase in $\sigma_r$ were observed and the wave energies were dissipated right after the shear. From Panel (a1), we can see that inside and shortly after the shear region, no switchbacks are observed, implying a strong dissipation of the wave energies. These results are consistent with the 2D MHD simulations \citep{Robertsetal1992,Shietal2020}, which showed that near the fast-slow stream interaction region, the wave energy is dissipated quickly because the shear transfers energies from long wavelengths to short wavelengths rapidly. They also found that inside the stream interaction region, the outward wave dominance is destroyed and kinetic energy exceeds the magnetic energy at small scales, which is consistent with the drop in $\sigma_c$ and increase in $\sigma_r$ observed by PSP. The positive $\sigma_r$ indicates that the velocity shear efficiently transfers kinetic energies from large to small scales. Thus, the velocity shear may play an important role in the turbulence evolution and is a good candidate to explain the observed negative $\sigma_c-R$ relation and positive $\sigma_r-R$ relation as discussed in Section \ref{sec:evolution_spectra}. 

\subsection{Non(low)-Alfv\'enic turbulence}
The middle column of Fig. \ref{fig:blow_up} shows the time period from 12:00 June 5 to 12:00 June 6, 2020 during Encounter 5. During this time period, and for most of Encounter 5 shown in Fig. \ref{fig:overview}, the turbulence property is ``abnormal.'' From Panel (a2), we can see that the magnetic field strength $|B|$ is quite constant and a lot of switchbacks are present. In addition, Panel (c2)\&(d2) show that the plasma density is quite constant with very small fluctuations. These features normally indicate a highly Alfv\'enic status of the turbulence. However, we can see from Panel (f2) that $\sigma_c$ is systematically small, around 0.5, and as is $\sigma_r$, which is around -0.75. That is to say, in this time period, there is a non-negligible amount of inward propagating wave component while magnetic energy significantly exceeds the kinetic energy, despite the near incompressibility. One can see from Fig. \ref{fig:overview} that actually during most of Encounter 5, the turbulence has low Alfv\'enicity and the wind speed is slow. 
In examining the middle column of Fig. \ref{fig:overview}, we noticed that in Encounter 4 after the heliospheric current sheet crossing on February 1 until February 4, the solar wind was also quite slow and had relatively low $\sigma_c$ and $\sigma_r$, which is similar to what PSP observed in Encounter 5. Thus, the observed non-Alfv\'enic, or low-Alfv\'enic, turbulence is possibly related with the sources of the very slow solar wind. One thing that we should point out is that the ion density measured by the Faraday cup (SPC) seems to be lower than the electron density derived using the quasi thermal noise (QTN) measurements made by the Radio Frequency Spectrometer Low Frequency Receiver (RFS/LFR) \citep{moncuquet2020first}. In Fig. \ref{fig:compare_SPC_QTN}, we plotted these two quantities for Encounter 5, where blue is the SPC ion density $n_p$ and orange is the QTN electron density $n_e$. We can see that $n_p$ is systematically lower than $n_e$ and the difference can be as large as $\sim 30\%$ for some time periods. As we expect that the QTN measurements are more accurate than the SPC measurements, this indicates that the real ion density is larger than the SPC data used in the current study. As a result, the magnetic energy density $E_b = b^2/\mu_0 \rho$ calculated here is larger than real, leading to an overestimate of the magnetic energy excess over the kinetic energy. Thus, we used the QTN-derived density to reconduct the calculation of the magnetic energy, $\sigma_c$ and $\sigma_r$. The result is not presented here but we confirm that the effect of this density difference is not significant and does not change the low-Alfv\'enicity in E5.

In Fig. \ref{fig:connection_to_corona}, we show the SDO/HMI image of the whole disk of the Sun taken on June 16, 2020. During most of Encounter 5, PSP was flying over this side of the Sun, which is very quiet as can be seen from the image. We note that this image was not taken during the period that PSP data were analyzed (May 30-June 13, 2020) since PSP was not on the Sun-Earth line during E5 so there is a time lag between the encounter and when SDO was looking at the solar surface over which PSP flew by. In the right panel of Fig. \ref{fig:connection_to_corona}, we show the map of magnetic pressure at $R=1.2R_s$, which was calculated using the PFSS model with the source surface set to $R_{ss}= 2.5R_s$ and the SDO/HMI measurements as input. The blue diamond is the direct radial projection of PSP to the source surface and the blue crosses are the foot points of the magnetic field lines connected to PSP on the source surface. Different crosses correspond to a prediction using varying wind speeds, from 230-80km/s to 230+80km/s. The blue circles are on the surface $R=1.2R_s$ and are magnetically connected to the blue crosses according to the PFSS model results. The detailed procedure to create this plot can be found in \citet{panasenco2020exploring} and Velli et al. (2021, this issue). We can see that at this time period PSP was connected to the boundary of the northern polar coronal hole without any activities nearby, neither active regions, nor pseudo-streamers, which are shown to be crucial in generating the Alfv\'enic slow wind observed in Encounter 1 \citep{panasenco2020exploring}. For most of E5, PSP was magnetically connected to the boundaries of either the northern or southern polar coronal hole (Velli et al., 2021, this issue). This may be relevant to explain why the slow wind observed during Encounter 5 is non-Alfv\'enic despite of the quite incompressible fluctuations. One possibility is the different ion compositions in the slow wind originating from different regions. For example, if the slow wind that originates near the boundaries of polar coronal holes comprises more helium or heavier ions which are not considered in the current study, the real plasma density should be larger than our estimate. As a result, the real magnetic energy density should be smaller than our calculation. If so, $\sigma_r$ should be closer to 0 and $\sigma_c$ should be closer to 1, that is the Alfv\'enicity of the wind should be larger than our estimate. Further analysis of the ion composition is necessary, but this is beyond the scope of the current study. Other mechanisms are also possible. For example, if the Alfv\'en waves in the slow wind originating near the polar coronal holes experience strong reflection due to large inhomogeneity of the background Alfv\'en velocity, the Alfv\'enicity is low. Modeling the propagation of Alfv\'en waves at different regions of the Sun will be a future topic. We conclude here that the coronal magnetic structures play a key role in the Alfv\'enic properties of the solar wind.

\subsection{Effect of the heliospheric current sheets}

The right column of Fig. \ref{fig:blow_up} shows the time period from 18:00 June 7 to 00:00 June 9, 2020 during Encounter 5. In this time period, PSP crossed a plasma sheet, inside which the ion density, speed, and temperature were all enhanced while the magnetic field strength was weakened with multiple polarity reversals. These measurements imply that PSP crossed the heliospheric current sheet, which is typically embedded inside a plasma sheet \citep{Smith2001}, multiple times. The turbulence properties inside this plasma sheet are very different compared with those in the normal solar wind streams. First, the spectra of both the magnetic field and velocity become steeper, with slopes close to $-2$ because of the frequent discontinuities. Second, $\sigma_c$ is on average close to 0, that is there are no well-defined Alfv\'enic fluctuations or the outward and inward propagating Alfv\'en waves are strongly mixed. Third, $\sigma_r$ is close to -1, implying magnetic-dominant fluctuations. During Encounter 4, from January 17 to January 20, 2020, PSP also crossed current sheets multiple times and one can observe from the middle column of Fig. \ref{fig:overview} that in this time period, $\sigma_c$ was frequently negative and $\sigma_r$ was very low. These measurements suggest that current sheets may also play an important role in generating the low $\sigma_c$ and $\sigma_r$ fluctuations observed in the slow streams such as that shown in the middle column of Fig. \ref{fig:blow_up}. \citet{malara1996gompressive}, via 2.5D MHD simulations of Alfv\'en waves on top of a current sheet, showed that the initially large $\sigma_c$ is rapidly destroyed in the vicinity of the current sheet, supporting our observation. 
In assuming that these fluctuations in the slow wind are strongly affected by current sheets such that they are non-Alfv\'enic at their origins, then we need to explain why the magnitude of magnetic field is still nearly constant. Firehose instability may play a key role in explaining this as \citet{tenerani2018nonlinear} show that magnetic field fluctuations in high-$\beta$ plasma naturally relax to a constant-$|B|$ status due to the firehose instability.

\section{Conclusions}
In this study, we have analyzed data from the first five orbits of PSP. We focus on the properties of the MHD-scale turbulence and how they vary with the large-scale solar wind streams. A general nonlinear steepening of the magnetic field spectrum from a $-3/2$ slope toward $-5/3$ slope is observed statistically. The progress of the steepening depends on both the wind speed and the radial distance to the Sun, suggesting the existence of a ``turbulence age'' that controls the steepening process (see Fig. \ref{fig:spectral_slopes_2D}). The slope of velocity spectrum, on the contrary, remains almost constant at $-3/2$. The observed spectral evolution indicates that, on average, the magnetic field and velocity have similar spectra in the very young solar wind and their spectra evolve differently. Better theoretical models are still needed to explain this differential evolution of velocity and the magnetic field and they will be a future research topic. We investigated the Alfv\'enicity of the turbulence through two widely used diagnostics, namely the normalized cross helicity $\sigma_c$, which measures the relative abundance of outward and inward propagating Alfv\'en wave energies, and the normalized residual energy $\sigma_r$, which measures the relative abundance of magnetic and kinetic energies. Statistically, turbulence in fast solar wind is more ``Alfv\'enic'' than that in slow wind as $\sigma_c$ is closer to 1 and $\sigma_r$ is closer to 0 in the fast wind. During radial evolution, in general, the dominance of an outward propagating wave gradually weakens, manifested in a decreasing $\sigma_c$ (see left panel of Fig. \ref{fig:sigma_c_sigma_r}). The magnetic-kinetic energy comparison is surprising as our result shows that the magnetic energy significantly exceeds the kinetic energy close to the Sun and gradually relaxes to a balanced status. This is in contrast to the commonly accepted idea that the magnetic energy excess is a result of the dynamic evolution of MHD turbulence \citep[e.g.,][]{grappin1983dependence}. A similar result was reported by \citet{bavassano1998cross}, who analyzed Ulysses data and showed that the least evolved high-latitude stream has the strongest imbalance between magnetic and kinetic energies compared with more evolved mid- and low-latitude streams. They attributed this phenomenon to the abundance of pickup ions in the polar region, which modifies the kinetic normalization of the Alfv\'enic unit. However, other mechanisms, such as the contribution of heavy ions and the effect of the velocity shears, may also play important roles.

We note that the above results are all based on a statistical analysis. In practice, individual streams can be quite different from each other and one cannot simply infer the turbulence properties from the wind speed. For example, from Fig. \ref{fig:overview} \& \ref{fig:blow_up}, we observe that the slow streams with a similar speed ($\sim 300$km/s) can be either highly Alfv\'enic (Encounter 1) or non-Alfv\'enic (Encounters 4\&5). To fully understand the cause of these differences, we must examine the origin of each individual solar wind stream because the location of the origin can significantly impact the Alfv\'enicity of the slow wind \citep{DAmicisandBruno2015,panasenco2020exploring}. In addition, it is possible that the large-scale structures, such as the heliospheric current sheets and velocity shears, greatly modify the turbulence properties at the very early stage \citep[e.g.,][]{Robertsetal1992,Shietal2020}.

\begin{acknowledgements}
This research was funded in part by the FIELDS experiment on the Parker Solar Probe spacecraft, designed and developed under NASA contract NNN06AA01C and the NASA Parker Solar Probe Observatory Scientist grant NNX15AF34G.
\end{acknowledgements}

%
  \bibliographystyle{aa} 
  \bibliography{references} 
%



\end{document}